\documentclass[fleqn,twoside]{article}
\usepackage{espcrc2}

\usepackage{graphicx}
\usepackage[figuresright]{rotating}

\newcommand{\AmS}{{\protect\the\textfont2
  A\kern-.1667em\lower.5ex\hbox{M}\kern-.125emS}}
\title{Photon Total Cross-sections\thanks{Supported in part by 
EEC RTN-CT2002-311 and
by the Department of Science and
Technology, India, project number SP/S2/K-01/2000-II.}}

\author{R.M. Godbole \address{
Centre for Theoretical Studies,
Indian Institute of Science,
Bangalore 560012, India }, A. Grau \address{Departamento
 de F\'\i sica Te\'orica y del Cosmos, University of Granada,
     18071   Granada, Spain},
G. Pancheri \address{INFN, Frascati National Laboratories, I00044 Frascati,
  Italy} 
and Y.N. Srivastava \address{INFN and Physics Department, University of
  Perugia, I06123 Perugia, Italy}.}
       
\begin{document}
\begin{titlepage}
\begin{flushright}
                                                   IISc-CTS/10/03\\
                                   LNF-03/20 (P), 17 Novembre 2003\\
                                                   hep-ph/0311211  \\
\end{flushright}
                                                                                
\vspace{0.5cm}
\begin{center}
{\Large  
{\bf  Photon Total Cross-sections }}\\[5ex]
R.M. Godbole$^{a}$ , A. Grau$^{b}$,  G. Pancheri$^{c}$ and Y.N. Srivastava$^{d}$
\\[2ex]
a: {\it Centre for Theoretical Studies, Indian Institute of Science, Bangalore,
560 012, India,}\\[1.5ex]
b:{\it Departamento de F\'\i sica Te\'orica y del Cosmos, University of Granada,
     18071   Granada, Spain,}\\[1.5ex]
c: {\it INFN, Frascati National Laboratories, I00044 Frascati, Italy,}\\[1.5ex]
d: {\it INFN and Physics Department, University of Perugia, I06123 Perugia, 
Italy.}\\
\end{center}
\vspace{1.0cm}
{\begin{center}

ABSTRACT
                                                                                
\vspace{1cm}                                                                                
\parbox{15cm}{
We discuss present predictions for the total $\gamma \gamma$ and $\gamma p
$ cross-sections, highlighting why predictions differ. We present results
from the Eikonal Minijet Model and  improved predictions based on
soft gluon resummation.
}
\end{center}}

\vfill
{\begin{center}
{\it Talk presented  by G. Pancheri at}                       \\
{\it PHOTON-2003, International Meeting on Structure and Interactions of the Photon} \\
{\it Frascati, Italy, April 7-11, 2003}
\end{center}}
                                                                                
\vfill
\end{titlepage}

\begin{abstract}
We discuss present predictions for the total $\gamma \gamma$ and $\gamma p
$ cross-sections, highlighting why predictions differ. We present results
from the Eikonal Minijet Model and  improved predictions based on
soft gluon resummation. 
\vspace{1pc}
\end{abstract}

\maketitle

\section{Present predictions for $\gamma \gamma\rightarrow$ hadrons}
\begin{figure}[htb]
\includegraphics[width=17pc]
{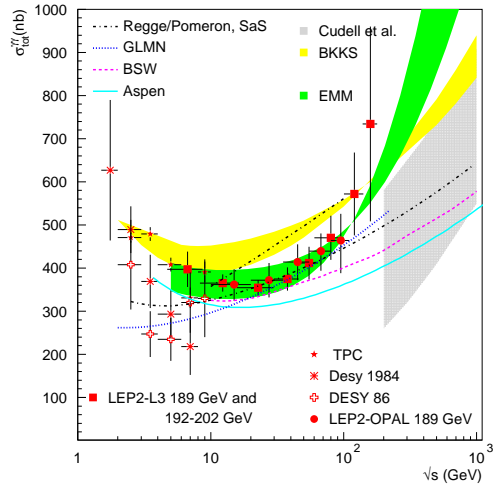}
\caption{Predictions for $\gamma \gamma \rightarrow hadrons$ from 
various models, Aspen \cite{aspen}, SaS\cite{SaS}, BSW\cite{BSW},
GLMN \cite{GLMN}, BKKS \cite{BKKS}, EMM \cite{EMM,epjc} and Cudell 
et al. \cite{cudell}.}
\label{fig1}
\end{figure}

\begin{figure}[htb]
\includegraphics[width=17pc]
{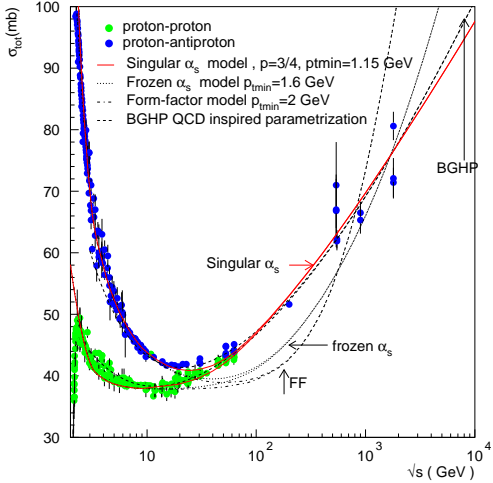}
\caption{Total proton proton and proton-antiproton cros-section as
  described by the Aspen model \cite{aspen}(labelled BHGP), and
  a QCD mini-jet model which includes soft gluon effects\cite{ppbn}. 
Tevatron data come from E710\cite{E710}, E811\cite{E811} and
  CDF \cite{CDF} experiments.}
\label{fig2}
\end{figure}

\begin{figure}[htb]
\includegraphics[width=17pc]
{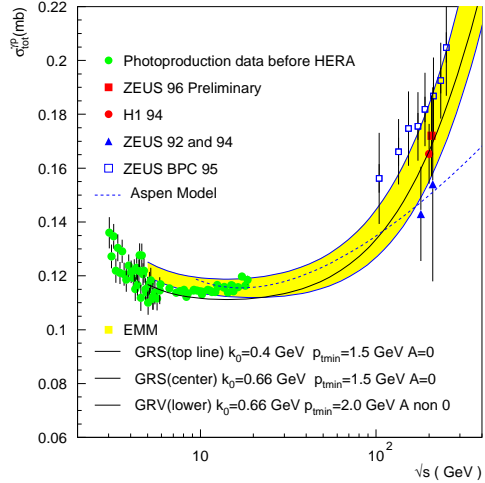}
\caption{Data for total $\gamma p \rightarrow hadrons$ and predictions
  from the Aspen\cite{aspen} and EMM\cite{EMM,epjc} model. HERA data are from
  ZEUS
\cite{ZEUS}, H1 \cite{H1} and a set of data extrapolated
 from $Q^2\neq 0$ from the ZEUS BPC \cite{BPC}.}
\label{fig3}
\end{figure}

Present predictions of $\gamma \gamma \rightarrow hadrons$ 
at energies covered by the Linear Collider differ by  large factors\cite{jhep},
 as we
show in Fig.\ref{fig1}. At $\sqrt{s}=500\ GeV$  different 
models can predict values
which differ by a factor 3, and the differences widen as the 
energy increase.
We plan, in the following, to discuss a  work program to reach
stable 
predictions, based on a QCD description of the decrease and   the 
rise of total cross-sections through Soft Gluon Summation
 (Bloch-Nordsieck Model)
and Mini-jets.
There are different reasons why predictions differ so widely one from the
other, some of which are related to the fact that there is no calculation 
to obtain quantitative descriptions of total cross-sections from first 
principles. This would 
    not necessarily be a deterrent from making correct predictions, as the
    $pp/p{\bar p}$ case shows. In Fig. \ref{fig2} we show present data and some
    model predictions for the proton case.
Another important reason for the variety of predictions is that 
all models for $\gamma \gamma$ apply some degree of extrapolation 
from $\gamma p
$ and $pp/p{\bar p}$ data. Since, for both photon and proton processes,
 there are still
 differences among data at high
  energy (although within one or two standard deviations at most)
this ends up  doubling the errors in the extrapolation to
  $\gamma \gamma$.  
 The present range of
  variability of the high energy data for the photoproduction cross-section
is highlighted in Fig. \ref{fig3}, where present data are shown 
together with the predictions from the
  Eikonal Minijet Model (EMM)\cite{epjc}.

As for $\gamma \gamma$, 
it should also be pointed out that at low energies old 
$\gamma \gamma$ data have large errors and even
  LEP data \cite{martinkhang} may   have a 10\% normalization error. 
Finally, $\gamma \gamma$ data 
 do not reach a high enough energy to pinpoint how the
  cross-section rises (unlike  the $pp/p{\bar p}$ case). These reasons make
widely varying   predictions for $\gamma \gamma \rightarrow hadrons$.

\section{Which predictions to trust}

We can distinguish between various models by grouping them as those for which 
 the photon is treated like a proton  vs. the QCD models. 
To the first group there
 belong also models based on Gribov factorization
\begin{equation}
\sigma_{\gamma \gamma}={{\sigma_{\gamma p}^2}\over{\sigma_{pp/{\bar p}}}}
\end{equation}
for which $ 
\sigma_{\gamma \gamma}(\sqrt{s}=1\ TeV)=500\div 700\ nb$.The QCD based
models include the Eikonal Minijet Model (EMM) for which $
\sigma_{\gamma \gamma}(\sqrt{s}=1\ TeV)=1000\div 1500\ nb$.
We show in Fig. 4 two different 
 predictions from the EMM,  which will be discussed 
shortly.
\begin{figure}
\includegraphics[width=17pc]
{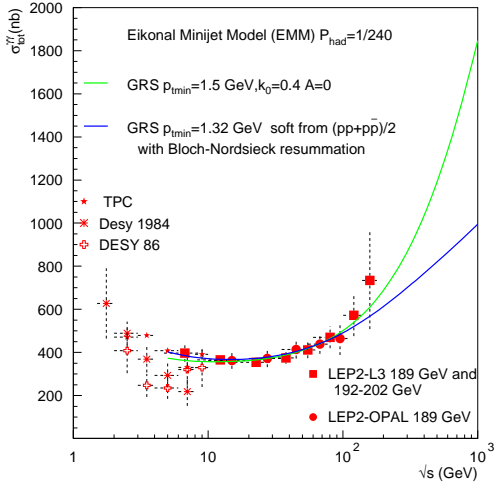}
\caption{Data for $\gamma \gamma \rightarrow hadrons$ and fits using the
  EMM, with and without soft gluon resumation. }

\label{fig4}
\end{figure}
\section{QCD vs. stable predictions}
A work program to reach stable predictions will be based on
treating the photon at low energy like a proton, while distinguishing it
from the proton at high energy where QCD processes and parton densities may
be different for protons and photons. At the same time it will be important
to attempt a unified description for all three processes.
The basic expression for the total hadronic cross-section,
to be used throughout this paper, will be
based on the eikonal approximation, namely
\begin{equation}
\label{total}
\sigma_{tot}^{\gamma h}=2 P^{\gamma h}_{had} \int d^2{\vec b} 
[1-e^{-\chi_I(b,s)}\cos \chi_R(b,s)]
\end{equation}
where $P^{\gamma h}_{had}$ is a phenomenological parameter introduced to
describe the probability that a photon behaves like a
hadron. Its
value can be fixed from Quark Counting rules and Vector Meson
Dominance, to be
$P^{\gamma p}_{had}= \sum (4\pi \alpha /f^2_V)$,
$V=\rho,\omega,\phi$. For $\gamma \gamma$ processes Eq. \ref{total}
holds with $P^{\gamma \gamma}_{had}=[P^{\gamma p}_{had}]^2$. Eq. \ref{total}
is also used  for purely hadronic processes, in which case 
$P^{\gamma h}_{had}=1$.
We set $\chi_R(b,s)=0$  and from the expression for the 
inelastic cross-section, i.e.
\begin{equation}
\label{inel}
\sigma_{inel}^{\gamma h}=P^{\gamma h}_{had} \int d^2{\vec b} 
[1-e^{-2\chi_I(b,s)}]
\end{equation}
we identify $2\chi_I(b,s)$ with the average number $n(b,s)$ of 
inelastic collisions
taking place for any given value of the impact parameter $b$, at energy 
$\sqrt{s}$ of the colliding hadrons.
In the figures to follow, for all the curves 
with Bloch-Nordsieck
resummation, for $\gamma p$ we have chosen the soft
part of $n(b,s)$ as coming only from proton proton, as this 
 seems to give the
best description for the soft part, whereas for $\gamma \gamma$ we have
chosen
the average between $p p $ and $p {\bar p}$. Then, for $\gamma
\gamma$
 \begin{equation}
n(b,s)=n^{\gamma \gamma}_{soft} + 
A_{BN}(b,s,p_{tmin}){\tilde \sigma}^{\gamma \gamma}_{jets}(s,p_{tmin})
\end{equation}
with ${\tilde \sigma}^{\gamma \gamma}_{jets}(s,p_{tmin})=
 \sigma^{\gamma \gamma}_{jets}(s,p_{tmin})/P^{\gamma \gamma}_{had}$.
Resummation of soft gluons takes place 
through the Fourier transform of
the  exponentiated 
soft gluon transverse momentum distribution in $b$ space, obtained
 using the Bloch-Nordsieck (BN) method\cite{ppbn}, $
 e^{-h(b,s,p_{tmin})}$, with
\begin{equation}
h(b,s,p_{tmin})=\int_{k_{min}}^{k_{max}} d^3{\bar n}_{gluons}(k)\ 
[1-e^{ik_t\cdot  b}]
\end{equation}

In the BN model, the 
impact parameter space distribution appearing in the eikonal formalism is 
then identified with

\begin{equation}
A_{BN}(b,s,p_{tmin})={{
e^{-h(b,s,p_{tmin})}
}\over{
\int d^2{\vec b}\ e^{-h(b,s,p_{tmin})}
}}
\end{equation}
In our work program, we   first obtain a good description of proton
data\cite{tobe}. This  allows to fix
the soft eikonal to be used together with QCD minijets and  resummation for
protons. We then try to get a good description of $\gamma p$ using the soft
eikonal, 
and, subsequently, fix the jet  parameters,
 $p_{tmin}$ and densities, to be used with 
photons.
\section{Bloch-Nordsieck resummation}
Resummation and its embodyment in the EMM  constitute a very challenging
 task : this involves calculating the function
$h(b,s,p_{tmin})$, i.e. fix $k_{min}$ and 
$k_{max}$ for each parton parton scattering. In our presently simplified
 approach, we shall
   average the function $A_{BN}(b,s,p_{tmin})$, and hence $k_{max}$,
 over  densities and parton cross-sections, obtaining for $k_{max}$ a rising
 function of the energy $\sqrt{s}$, as discussed in the next section. 
A second crucial point of the BN approach, comes in setting $k_{min}=0$.
This
 requires the knowledge of 
$\alpha_s(k_t)$ as $\ k_t\rightarrow 0$\cite{yogi}. We use here
the model   in \cite{nak}, with an   ${\tilde \alpha}_s$ 
 singular but integrable
as discussed in \cite{yogi}, 
and such that for $k_\perp \gg \Lambda_{QCD}\ \ \ 
\ {\tilde \alpha}_s \to \alpha_s^{AF}$, while  for 
$k_\perp \ll \Lambda_{QCD}\ \ \ \  {\tilde \alpha}_s \to
(k^2_\perp)^{-p}$. Notice that
if p is smaller than 1 the integral in the function $h(b,s,p_{tmin})$ 
can be done.
\section{Energy dependence in impact parameter b}
To leading order in $\alpha_s$ the energy dependence which ultimately will 
soften the rise due to
mini-jets, comes from the maximum transverse momentum 
allowed to a single gluon emitted by the most energetic partons at the
beginning of the QCD cascade, valence quarks for the proton, all type of
quarks for the photon. The kinematics for the emission \cite{mario} gives
\begin{equation}
\label{qmaxjets}
k_{max}({\hat s})={{ \sqrt{{\hat s}} }\over{2}}
 (1-{{{\hat s_{jet}}}\over{{\hat s}}})
\end{equation}
with  integration to be done over 
 ${\hat s}$, the energy of the initial parton-parton subprocess
and the jet-jet invariant mass 
$\sqrt{{\hat s_{jet}}}$. Averaging over  densities
\begin{eqnarray}
 &  <k_{max}(s)>= 
{{\sqrt{s}} \over{2}}  \cdot \nonumber \\ 
&   \cdot {{ \sum_{\textstyle i,j}\int {{dx_1}\over{\sqrt{ x_1}}}
f_{i/a}(x_1)\int {{dx_2}\over{\sqrt{x_2}}}f_{j/b}(x_2) \int dz (1 - z)}
\over{\sum_{i,j}\int {dx_1\over x_1}
f_{i/a}(x_1)\int {{dx_2}\over{x_2}}f_{j/b}(x_2) \int(dz)}} 
  \nonumber
\end{eqnarray}
with 
    the lower limit of integration in the variable $z$ given by 
 $z_{min}=4p_{tmin}^2/(sx_1x_2)$.
\section{Soft Gluon Emission and  Energy Dependence}
The Bloch Nordsieck model is like the  EMM model with  
$\sigma_{jet}^{QCD} $ driving the rise. The Fourier transform of 
soft gluon emission  in $k_t$ space
gives  the impact parameter space distribution of
colliding partons. This  
introduces an  energy dependence in the  b-distribution of 
partons in the hadrons   which depends on 
 $p_{tmin}$ and the  parton densities. One achieves
two main results, a softening effect, and a reduction of  the
 dependence from  hard scattering parameters.
The softening effect happens 
because as $\sqrt{s}$ increases, the phase space available for
soft gluon emission also increases, and with it
 the transverse momentum of the initial colliding pair due to soft 
gluon emission. This leads to
 more straggling of initial partons and hence to  a reduced   probability
  for the collision.

\section{Bloch-Nordsieck Model for $p-p$ and $p-{\bar p}$}
In the proton-proton and proton-antiproton fit with the  Bloch-Nordsieck 
(BN) model, for the average number of collisions, we now write 
\begin{equation}
n(b,s)=\sigma_{soft} A^{soft}_{BN}+\sigma_{jet}A^{jet}_{BN}
\end{equation}
where $A_{BN}^{soft}(b,s)$ is obtained using the BN ans\"atz,  with a
$k_{max}$ which becomes  constant after a slight initial rise.
Soft gluon emission has now  a twofold effect as
 the energy increases: with  $\sigma_{soft}$  constant or decreasing (as
 from Regge exchange) $\sigma_{soft} A^{soft}_{BN}$ will decrease,
whereas,  with $\sigma_{jet}$ increasing rapidly, 
$\sigma_{jet} A^{jet}_{BN}$ will still increase
 but not as much as without soft gluons.
A good description is obtained with a soft part given by
\begin{equation}
\sigma^{pp}_{soft}=\sigma_0=48mb
\end{equation}
and
\begin{equation}
\sigma^{p{\bar p}}_{soft}=
\sigma_0 (1+{{2}\over{\sqrt{s}}}) 
\end{equation}
\begin{figure}
\includegraphics[width=17pc]
{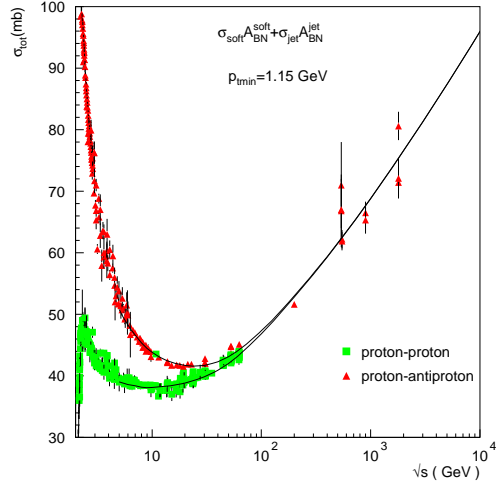}
\caption{Total proton-proton and proton-antiproton cross-section as
  described by the EMM with soft gluon emission both in the hard and soft 
region.}
\label{fig5}
\end{figure}

We show our present description \cite{tobe} of $pp$ and $p
{\bar p}$ data in Fig. \ref{fig5}.
\section{The case for $\gamma p$ and $\gamma \gamma$}
\begin{figure}
\includegraphics[width=17pc]
{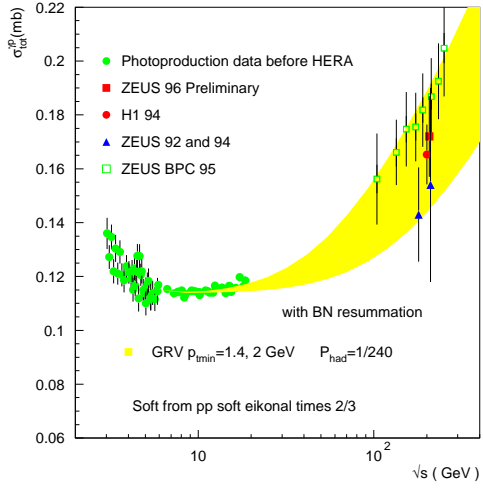}
\caption{Total $\gamma p$ cross-section, with soft gluon resummation (BN)
  and GRV densities in the
  mini-jet cross-section.}
\label{fig6}
\end{figure}

\begin{figure}
\includegraphics[width=17pc]
{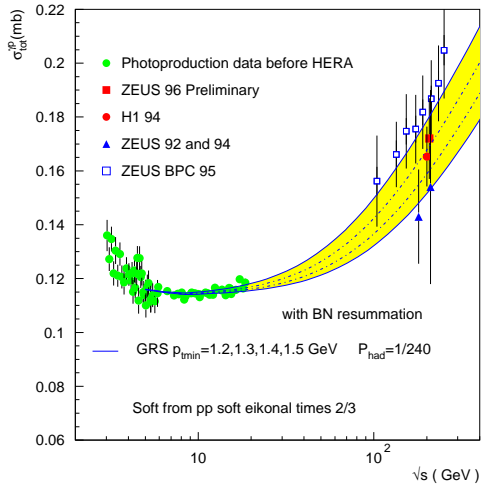}
\caption{Total $\gamma p$ cross-section, with soft gluon resummation (BN)
  and GRS densities in the
  mini-jet cross-section, for an indicative set of values for $p_{tmin}$.}
\label{fig7}
\end{figure}

\begin{figure}
\includegraphics[width=17pc]
{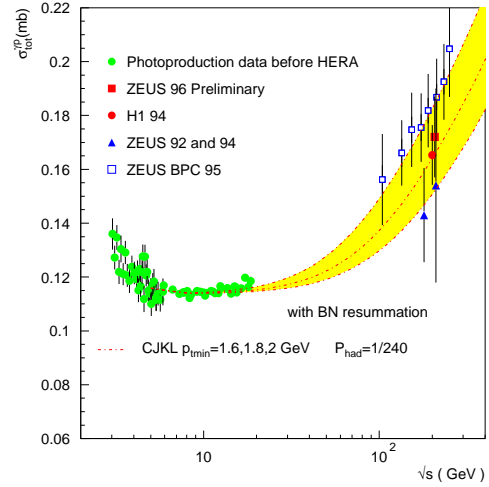}
\caption{As in Figs.\ref{fig6}, \ref{fig7} with CJKL densities and a set of
  $p_{tmin}$ values which fit the high energy data.}
\label{fig8}
\end{figure}

\begin{figure}
\includegraphics[width=17pc]
{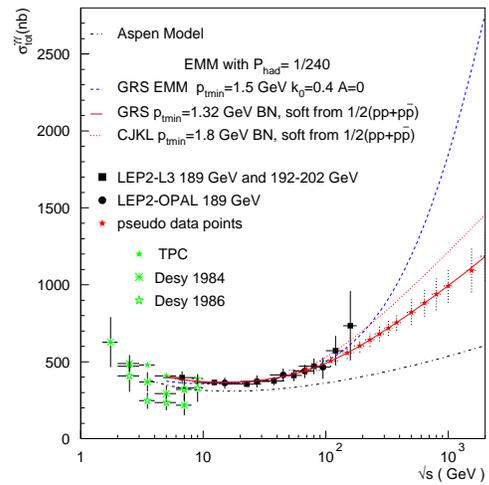}
\caption{Comparison between the Aspen model and the EMM with and without 
soft gluons, with   pseudo data points to be possibly  measured
 at the Linear Collider.}
\label{fig9}
\end{figure}

\begin{figure}

\includegraphics[width=17pc]
{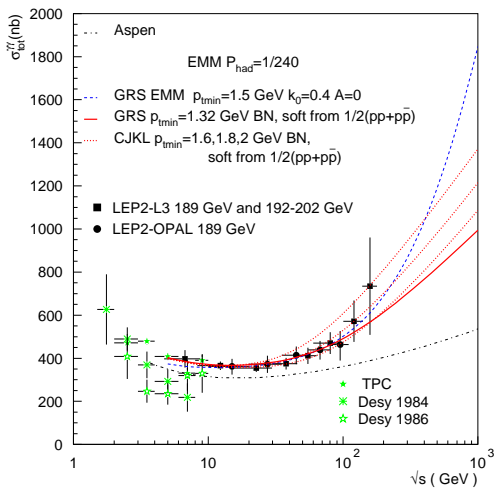}
\caption{Comparison between the Aspen model and the EMM with and without 
soft gluons, for different choices of parameters for the mini-jet 
cross-section.}
\label{fig10}
\end{figure}

With the previously described expressions, we now turn to $\gamma p$,
using $n^{\gamma p} _{soft}(b,s)={{2}\over{3}}n_{soft}^{pp}$. We  
 obtain various fits, depending upon the densities being used
for the photon, and 
the results are shown in Figs. \ref{fig6},\ref{fig7},\ref{fig8}, for each 
set of
densities and various values of $p_{tmin}$.
The present update for $\gamma \gamma$  is done using 
 the soft part of the eikonal $n(b,s)$  from the average of the proton 
and the antiproton fit, i.e.
 $n^{\gamma \gamma} _{soft}(b,s)={{4}\over{9}}(n_{soft}^{pp}+
n_{soft}^{p{\bar p}})/2$,
 soft resummation for hard scattering, and  three types of densities, 
GRV\cite{grv}, GRS\cite{grs} and CJKL \cite{cjlk}. In Fig. \ref{fig9}, we
 show a comparison between  the predictions from the Aspen model, the EMM
 without soft gluon emission, and two curves from the EMM with inclusion of
 soft gluons
 and different parton densities. We also indicate (stars) pseudo data
 points to be measured at the future Linear Collider.
How predictions for  $\gamma \gamma \rightarrow hadrons$ depend 
upon $p_{tmin}$ in the case of CJKL densities, can be seen in Fig. \ref{fig10}.
Similar results hold for other densities.
\section{Conclusions}
In this talk we have presented a comprehensive description of proton and photon 
total cross-sections, based on the use of the Eikonal representation and
on the hypothesis that QCD jet cross-sections
 drive the rise of all total cross-sections. This  Eikonal Minijet Model 
(EMM) is not fully satisfactory, since the rise with energy thus predicted is 
either too fast or too slow, depending on the parameters. 
It is shown that inclusion of soft gluon emission from initial state partons 
can give a much more realistic description in all cases, $pp, p{\bar p}, 
\gamma p$ and $\gamma \gamma$. Different models are also 
discussed and compared with the data and with the EMM.
\section*{Acknowledgements}
One of the authors, G.P., wishes to thank the MIT Center for Theoretical
Physics for hospitality during the writing of these Proceedings.

\end{document}